\documentstyle[prl,aps]{revtex}

\begin{document}
\draft

\title{A possible explanation for the anomalous acceleration of Pioneer 10}

\author{David F. Crawford}
\address{School of Physics, A28, University of Sydney, N.S.W. 2006, Australia\\
d.crawford@physics.usyd.edu.au}
\date{\today}

\maketitle

\begin{abstract}
The reported anomalous acceleration of the Pioneer 10 spacecraft of 
$\sim -8.5\times 10^{-10} \mbox{m.s}^{-2}$ (i.e. towards the sun) can be
explained by a gravitational interaction on the S-band signals 
traveling between Pioneer 10 and the earth. 
The effect of this gravitational interaction is 
a frequency shift that is proportional to the distance and the square root
of the density of the medium in which it travels.
If changes in this frequency are interpreted as a Doppler shift the result is 
an apparent acceleration directed towards the sun.
The gravitational interaction is caused by the focusing of the signal photons
in curved space where in this case the curvature is related to the density of the 
interplanetary dust.

\end{abstract}


\section{Introduction}
Precise tracking of the Pioneer 10/11, Galileo and Ulysses spacecraft
\cite{Anderson98} 
have shown an anomalous constant acceleration for Pioneer 10  with a magnitude 
$\sim -8.5\times 10^{-10}\, \mbox{m.s}^{-2}$. 
Additional analysis by the same team \cite{Turyshev99} provide  a new 
value for  the acceleration 
$(-7.5\pm 0.2)\times 10^{-10}\, \mbox{m.s}^{-2}$  
(where the uncertainty is estimated from points in their Fig 1) 
and  also reveal that there is an additional 
annual periodic component with a amplitude of  
$\sim 2\times 10^{-10}\, \mbox{m.s}^{-2}$ directed towards the sun.
The main method for monitoring the spacecraft is to measure the frequency 
shift of the signal returned by an active transponder. 
Any variation in this frequency shift that is not actually due to motion can be 
confused with
a Doppler shift and would be attributed to anomalous velocities 
and accelerations.

This paper argues that there are is an additional frequency shift in the 
spacecraft signal  
due to a gravitational interaction with the intervening material. 
Because the frequency shift is proportional to the distance to the spacecraft it
can easily mimic an acceleration.

\section{The explanation for the constant acceleration}
In previous papers \cite{Crawford79,Crawford87A,Crawford91} 
it was argued that photons have a 
gravitational interaction. This claim is based on the premise that in curved
space a bundle of geodesics is focused (the "focusing theorem", 
\cite{Misner73})
and as a consequence  photons are also focused. 
This leads to an interaction in which low energy photons are emitted and the primary photon losses energy. The effect can be observed as a frequency shift in a signal that is a function of distance traveled and the density of the local medium. 
Although the cosmological consequences of such an interaction are profound 
\cite{Crawford99}, it also leads to  predictions which can be tested locally,  
including the prediction \cite{Crawford91} that a 
frequency shift should be seen in the signals from spacecraft.
For a signal passing through a medium with matter density $\rho$ the rate of 
change of frequency, $f$, with distance is  \cite{Crawford87A,Crawford91}
\begin{equation}
\label{e1}
\frac{df}{dx} = -\left(\frac{8\pi G\rho}{c^{2}}\right)^{1/2}\!\! f.
\end{equation} 
Note that although point masses  may distort and deviate the geodesic 
bundle they do not focus it and so that there is no frequency shift 
predicted for signals passing near stars or planets. 
Since the effect is very small we can write it in effective velocity units as
\begin{equation}
\label{e2}
\Delta v = -\sqrt{8\pi G\rho}\,\Delta x .
\end{equation} 
Differentiating gives an apparent acceleration of $a = -\sqrt{8\pi G\rho}\,V$
where $V$ is the velocity of the spacecraft (or earth) and $\rho$ is the density at
the current positions.
It is not an average density over the path length.
Using  the observed anomalous acceleration of
$ -7.5\times 10^{-10}\, \mbox{m.s}^{-2}$  ,
and a Pioneer 10 velocity of $12.3 \mbox{km.s}^{-1}$, the 
required density for the two-way path is 
$5.5\times10^{-19}\,\mbox{kg.m}^{-3}$. 
The only constituent of the interplanetary medium that approaches this 
density is dust.
One estimate \cite{Sergeant80} of the interplanetary dust 
density at 1 AU is 
$1.3\times10^{-19}\,\mbox{kg.m}^{-3}$ and more recently Gr\"{u}n \cite{Grun99} 
suggests a value of
$10^{-19}\,\mbox{kg.m}^{-3}$ 
which is consistent with his earlier estimate of 
$9.6\times 10^{-20}\,\mbox{kg.m}^{-3}$ \cite{Grun85}. 
Although the authors do not give uncertainties it is clear that the densities could
be in error by a factor of two or more.
The main difficulties are the paucity of information and that the observations do
not span the complete range of grain sizes.
Taking a density of
$10^{-19}\,\mbox{kg.m}^{-3}$ 
the computed (anomalous) acceleration is 
$-3.4\times10^{-10}\,\mbox{ m.s}^{-2}$,
smaller by a factor of two than the observed anomalous acceleration.
However the density is required at the distance of Pioneer 10 in 1998 of 72 AU
in the plane of the ecliptic (ecliptic latitude of Pioneer 10 is 3$^\circ$). 

The meteroid experiment on-board Pioneer 10 measures the flux  of grains 
with masses larger than $10^{-10}\,$g.  
the results show  that after it left the influence
of Jupiter the flux \cite{Landgraf99}   was essentially constant 
(in fact there may be a slight rise) out to a distance of 18 AU.
It is thought that most of the grains are being continuously produced
in the Kuiper belt. As their orbits evolve inwards due to 
Poynting-Robertson drag and planetary perturbations they 
achieve a roughly constant spatial density.
Given the large  uncertainties in both the observed density at 1 AU (due to the
limitations of the detectors), and the extrapolation of the density to 72 AU, 
the conclusion is that interplanetary dust 
could provide the required density to explain the "anomalous acceleration"
by a frequency shift due to the gravitational interaction.

\section{The explanation for the annual acceleration}
Figure 1B in \cite{Turyshev99} shows a time varying acceleration 
that has a period
of one year and an amplitude that both fluctuates and decreases with time.
(It may not be a valid decrease but be due to the solar cycle.)
Their figure shows 50-day averages after the best-fit constant anomalous
acceleration has been removed.
For the years 1987 to 1993 where the curve is well defined the maxima occur at
$0.94\pm 0.03\,$yr and the minima at $0.45\pm 0.03\,$yr.
The amplitude changes from $\sim 2.5\times 10^{-10} \,\mbox{m.s}^{-2}$ in 1988 to
$\sim 1.5\times 10^{-10} \,\mbox{m.s}^{-2}$ in 1992.

In principle the gravitational interaction can explain this acceleration 
but now the relevant velocity is not that of Pioneer 10 but the 
orbital velocity of the earth.
Taking the earth's velocity as $30\,\mbox{km.s}^{-1}$ and a dust density of
$10^{-19}\,\mbox{kg.m}^{-3}$ 
the predicted annual acceleration in 1989 has an amplitude of
$7.6\times10^{-10}\,\mbox{ m.s}^{-2}$.
Although this acceleration is a factor of three too large a more
significant objection is that the predicted phase disagrees 
with the observations.
With this model the maximum accelerations should occur 
when the earth has a maximum velocity relative to Pioneer 10, 
namely when it has maximum elongation as seen from the spacecraft.
Since in 1989 Pioneer 10 had an ecliptic longitude of $\sim 72^\circ$ 
these should occur at 0.17\,yr and 0.68\,yr. 
The discrepancy in phase of $97^\circ \pm 11^\circ$ means that the gravitational 
interaction does not  directly explain the annual variation.
However since the gravitational interaction was not included in in the 
complex calculations used to 
compute the trajectory it is feasible that the effect has been 
compensated for by small adjustments to
other parameters and all that is left is a distorted residual.

If mistakenly interpreted as a Doppler shift the annual component of 
the gravitational interaction 
is equivalent to an additional velocity of the earth (as seen by Pioneer 10) of 
$3.8\,\mbox{mm.s}^{-1}$.
For a circular orbit of the earth  this is equivalent to a shift in the 
longitude of Pioneer 10 of 0.026 arcseconds.
Thus if there is a gravitational interaction it could be masked by a small error
in longitude.
In practice the position of Pioneer 10 must be consistent with celestial
mechanics and many other observations and it is unlikely that there would
be complete compensation.
The final analysis requires the inclusion of the gravitational interaction into
the orbit calculations.

\section{Conclusion}
It has been argued that the gravitational interaction with a
interplanetary dust density of
$10^{-19}\,\mbox{kg.m}^{-3}$ 
predicts an anomalous acceleration  of Pioneer 10 at 72 AU of
$-3.4\times10^{-10}\,\mbox{ m.s}^{-2}$
to be compared with the observed value of
$(-7.5\pm 0.2)\times 10^{-10}\, \mbox{m.s}^{-2}$.
The largest uncertainty is in the estimate of the interplanetary dust density.
Since the annual period in the gravitational interaction is easily masked by
small shift in the longitude of Pioneer 10 its
effects are unlikely to be observed.
However the predicted magnitude is in the right range  
and the observed annual acceleration  could be the residuals after
a partial compensation.

\section{Acknowledgments}
This work is supported by the Science Foundation for Physics within the 
University of Sydney, 
and use has made of NASA's Astrophysics Data System Abstract Service. 


\end{document}